\documentclass[twocolumn,letterpaper,aps,prc,longbibliography,superscriptaddress,nofootinbib,floatfix]{revtex4-2}

\usepackage{graphicx}	
\usepackage{amsmath}
\usepackage{xspace}	

\newcommand{\pt}{\mbox{$p_T$}\xspace}

\newcommand{\sqsn}{\mbox{$\sqrt{s_{_{NN}}}$}\xspace}
\newcommand{\sqsntwo}{\mbox{$\sqrt{s_{_{NN}}}=200$~GeV}\xspace}

\newcommand{\pp}{\mbox{$p$$+$$p$}\xspace}
\newcommand{\dau}{\mbox{$d$$+$Au}\xspace}

\newcommand{\pdhe}{\mbox{$p/d/^3$He}\xspace}
\newcommand{\pau}{\mbox{$p$$+$Au}\xspace}

\newcommand{\vtwo}{\mbox{$v_2$}\xspace}
\newcommand{\vthree}{\mbox{$v_3$}\xspace}
\newcommand{\vn}{\mbox{$v_n$}\xspace}

\newcommand{\ppb}{\mbox{$p$$+$Pb}\xspace}
\newcommand{\heau}{\mbox{$^3$He$+$Au}\xspace}

\begin{document}


\title{Measurements of second-harmonic Fourier coefficients from azimuthal 
anisotropies in $p$$+$$p$, $p$$+$Au, $d$$+$Au, and $^3$He$+$Au collisions 
at $\sqrt{s_{_{NN}}}=200$ GeV}

\newcommand{\abilene}{Abilene Christian University, Abilene, Texas 79699, USA}
\newcommand{\augie}{Department of Physics, Augustana University, Sioux Falls, South Dakota 57197, USA}
\newcommand{\banaras}{Department of Physics, Banaras Hindu University, Varanasi 221005, India}
\newcommand{\barc}{Bhabha Atomic Research Centre, Bombay 400 085, India}
\newcommand{\baruch}{Baruch College, City University of New York, New York, New York, 10010 USA}
\newcommand{\bnlcoll}{Collider-Accelerator Department, Brookhaven National Laboratory, Upton, New York 11973-5000, USA}
\newcommand{\bnlphys}{Physics Department, Brookhaven National Laboratory, Upton, New York 11973-5000, USA}
\newcommand{\caucr}{University of California-Riverside, Riverside, California 92521, USA}
\newcommand{\charlesczech}{Charles University, Faculty of Mathematics and Physics, 180 00 Troja, Prague, Czech Republic}
\newcommand{\ciae}{Science and Technology on Nuclear Data Laboratory, China Institute of Atomic Energy, Beijing 102413, People's Republic of China}
\newcommand{\cns}{Center for Nuclear Study, Graduate School of Science, University of Tokyo, 7-3-1 Hongo, Bunkyo, Tokyo 113-0033, Japan}
\newcommand{\colorado}{University of Colorado, Boulder, Colorado 80309, USA}
\newcommand{\columbia}{Columbia University, New York, New York 10027 and Nevis Laboratories, Irvington, New York 10533, USA}
\newcommand{\czechtech}{Czech Technical University, Zikova 4, 166 36 Prague 6, Czech Republic}
\newcommand{\debrecen}{Debrecen University, H-4010 Debrecen, Egyetem t{\'e}r 1, Hungary}
\newcommand{\elte}{ELTE, E{\"o}tv{\"o}s Lor{\'a}nd University, H-1117 Budapest, P{\'a}zm{\'a}ny P.~s.~1/A, Hungary}
\newcommand{\ewha}{Ewha Womans University, Seoul 120-750, Korea}
\newcommand{\famu}{Florida A\&M University, Tallahassee, FL 32307, USA}
\newcommand{\fsu}{Florida State University, Tallahassee, Florida 32306, USA}
\newcommand{\gsu}{Georgia State University, Atlanta, Georgia 30303, USA}
\newcommand{\hiroshima}{Hiroshima University, Kagamiyama, Higashi-Hiroshima 739-8526, Japan}
\newcommand{\howard}{Department of Physics and Astronomy, Howard University, Washington, DC 20059, USA}
\newcommand{\ihepprot}{IHEP Protvino, State Research Center of Russian Federation, Institute for High Energy Physics, Protvino, 142281, Russia}
\newcommand{\illuiuc}{University of Illinois at Urbana-Champaign, Urbana, Illinois 61801, USA}
\newcommand{\inrras}{Institute for Nuclear Research of the Russian Academy of Sciences, prospekt 60-letiya Oktyabrya 7a, Moscow 117312, Russia}
\newcommand{\instpasczech}{Institute of Physics, Academy of Sciences of the Czech Republic, Na Slovance 2, 182 21 Prague 8, Czech Republic}
\newcommand{\isu}{Iowa State University, Ames, Iowa 50011, USA}
\newcommand{\jaea}{Advanced Science Research Center, Japan Atomic Energy Agency, 2-4 Shirakata Shirane, Tokai-mura, Naka-gun, Ibaraki-ken 319-1195, Japan}
\newcommand{\jeonbuk}{Jeonbuk National University, Jeonju, 54896, Korea}
\newcommand{\jyvaskyla}{Helsinki Institute of Physics and University of Jyv{\"a}skyl{\"a}, P.O.Box 35, FI-40014 Jyv{\"a}skyl{\"a}, Finland}
\newcommand{\kek}{KEK, High Energy Accelerator Research Organization, Tsukuba, Ibaraki 305-0801, Japan}
\newcommand{\korea}{Korea University, Seoul 02841, Korea}
\newcommand{\kurchatov}{National Research Center ``Kurchatov Institute", Moscow, 123098 Russia}
\newcommand{\kyoto}{Kyoto University, Kyoto 606-8502, Japan}
\newcommand{\lawllnl}{Lawrence Livermore National Laboratory, Livermore, California 94550, USA}
\newcommand{\losalamos}{Los Alamos National Laboratory, Los Alamos, New Mexico 87545, USA}
\newcommand{\lund}{Department of Physics, Lund University, Box 118, SE-221 00 Lund, Sweden}
\newcommand{\lyon}{IPNL, CNRS/IN2P3, Univ Lyon, Universit{\'e} Lyon 1, F-69622, Villeurbanne, France}
\newcommand{\maryland}{University of Maryland, College Park, Maryland 20742, USA}
\newcommand{\mass}{Department of Physics, University of Massachusetts, Amherst, Massachusetts 01003-9337, USA}
\newcommand{\mate}{MATE, Laboratory of Femtoscopy, K\'aroly R\'obert Campus, H-3200 Gy\"ongy\"os, M\'atrai\'ut 36, Hungary}
\newcommand{\michigan}{Department of Physics, University of Michigan, Ann Arbor, Michigan 48109-1040, USA}
\newcommand{\miss}{Mississippi State University, Mississippi State, Mississippi 39762, USA}
\newcommand{\muhlenberg}{Muhlenberg College, Allentown, Pennsylvania 18104-5586, USA}
\newcommand{\nara}{Nara Women's University, Kita-uoya Nishi-machi Nara 630-8506, Japan}
\newcommand{\natmephi}{National Research Nuclear University, MEPhI, Moscow Engineering Physics Institute, Moscow, 115409, Russia}
\newcommand{\newmex}{University of New Mexico, Albuquerque, New Mexico 87131, USA}
\newcommand{\nmsu}{New Mexico State University, Las Cruces, New Mexico 88003, USA}
\newcommand{\northcg}{Physics and Astronomy Department, University of North Carolina at Greensboro, Greensboro, North Carolina 27412, USA}
\newcommand{\ohio}{Department of Physics and Astronomy, Ohio University, Athens, Ohio 45701, USA}
\newcommand{\ornl}{Oak Ridge National Laboratory, Oak Ridge, Tennessee 37831, USA}
\newcommand{\orsay}{IPN-Orsay, Univ.~Paris-Sud, CNRS/IN2P3, Universit\'e Paris-Saclay, BP1, F-91406, Orsay, France}
\newcommand{\peking}{Peking University, Beijing 100871, People's Republic of China}
\newcommand{\pnpi}{PNPI, Petersburg Nuclear Physics Institute, Gatchina, Leningrad region, 188300, Russia}
\newcommand{\pusan}{Pusan National University, Pusan 46241, Korea}
\newcommand{\riken}{RIKEN Nishina Center for Accelerator-Based Science, Wako, Saitama 351-0198, Japan}
\newcommand{\rikjrbrc}{RIKEN BNL Research Center, Brookhaven National Laboratory, Upton, New York 11973-5000, USA}
\newcommand{\rikkyo}{Physics Department, Rikkyo University, 3-34-1 Nishi-Ikebukuro, Toshima, Tokyo 171-8501, Japan}
\newcommand{\saispbstu}{Saint Petersburg State Polytechnic University, St.~Petersburg, 195251 Russia}
\newcommand{\seoulnat}{Department of Physics and Astronomy, Seoul National University, Seoul 151-742, Korea}
\newcommand{\stonybrkc}{Chemistry Department, Stony Brook University, SUNY, Stony Brook, New York 11794-3400, USA}
\newcommand{\stonycrkp}{Department of Physics and Astronomy, Stony Brook University, SUNY, Stony Brook, New York 11794-3800, USA}
\newcommand{\tenn}{University of Tennessee, Knoxville, Tennessee 37996, USA}
\newcommand{\texsu}{Texas Southern University, Houston, TX 77004, USA}
\newcommand{\titech}{Department of Physics, Tokyo Institute of Technology, Oh-okayama, Meguro, Tokyo 152-8551, Japan}
\newcommand{\tsukuba}{Tomonaga Center for the History of the Universe, University of Tsukuba, Tsukuba, Ibaraki 305, Japan}
\newcommand{\vandy}{Vanderbilt University, Nashville, Tennessee 37235, USA}
\newcommand{\weizmann}{Weizmann Institute, Rehovot 76100, Israel}
\newcommand{\wigner}{Institute for Particle and Nuclear Physics, Wigner Research Centre for Physics, Hungarian Academy of Sciences (Wigner RCP, RMKI) H-1525 Budapest 114, POBox 49, Budapest, Hungary}
\newcommand{\yonsei}{Yonsei University, IPAP, Seoul 120-749, Korea}
\newcommand{\zagreb}{Department of Physics, Faculty of Science, University of Zagreb, Bijeni\v{c}ka c.~32 HR-10002 Zagreb, Croatia}
\newcommand{\zambia}{Department of Physics, School of Natural Sciences, University of Zambia, Great East Road Campus, Box 32379, Lusaka, Zambia}
\affiliation{\abilene}
\affiliation{\augie}
\affiliation{\banaras}
\affiliation{\barc}
\affiliation{\baruch}
\affiliation{\bnlcoll}
\affiliation{\bnlphys}
\affiliation{\caucr}
\affiliation{\charlesczech}
\affiliation{\ciae}
\affiliation{\cns}
\affiliation{\colorado}
\affiliation{\columbia}
\affiliation{\czechtech}
\affiliation{\debrecen}
\affiliation{\elte}
\affiliation{\ewha}
\affiliation{\famu}
\affiliation{\fsu}
\affiliation{\gsu}
\affiliation{\hiroshima}
\affiliation{\howard}
\affiliation{\ihepprot}
\affiliation{\illuiuc}
\affiliation{\inrras}
\affiliation{\instpasczech}
\affiliation{\isu}
\affiliation{\jaea}
\affiliation{\jeonbuk}
\affiliation{\jyvaskyla}
\affiliation{\kek}
\affiliation{\korea}
\affiliation{\kurchatov}
\affiliation{\kyoto}
\affiliation{\lawllnl}
\affiliation{\losalamos}
\affiliation{\lund}
\affiliation{\lyon}
\affiliation{\maryland}
\affiliation{\mass}
\affiliation{\mate}
\affiliation{\michigan}
\affiliation{\miss}
\affiliation{\muhlenberg}
\affiliation{\nara}
\affiliation{\natmephi}
\affiliation{\newmex}
\affiliation{\nmsu}
\affiliation{\northcg}
\affiliation{\ohio}
\affiliation{\ornl}
\affiliation{\orsay}
\affiliation{\peking}
\affiliation{\pnpi}
\affiliation{\pusan}
\affiliation{\riken}
\affiliation{\rikjrbrc}
\affiliation{\rikkyo}
\affiliation{\saispbstu}
\affiliation{\seoulnat}
\affiliation{\stonybrkc}
\affiliation{\stonycrkp}
\affiliation{\tenn}
\affiliation{\texsu}
\affiliation{\titech}
\affiliation{\tsukuba}
\affiliation{\vandy}
\affiliation{\weizmann}
\affiliation{\wigner}
\affiliation{\yonsei}
\affiliation{\zagreb}
\affiliation{\zambia}
\author{N.J.~Abdulameer} \affiliation{\debrecen}
\author{U.~Acharya} \affiliation{\gsu} 
\author{A.~Adare} \affiliation{\colorado} 
\author{C.~Aidala} \affiliation{\michigan} 
\author{N.N.~Ajitanand} \altaffiliation{Deceased} \affiliation{\stonybrkc} 
\author{Y.~Akiba} \email[PHENIX Spokesperson: ]{akiba@rcf.rhic.bnl.gov} \affiliation{\riken} \affiliation{\rikjrbrc} 
\author{M.~Alfred} \affiliation{\howard} 
\author{V.~Andrieux} \affiliation{\michigan} 
\author{K.~Aoki} \affiliation{\kek} \affiliation{\riken} 
\author{N.~Apadula} \affiliation{\isu} \affiliation{\stonycrkp} 
% ...
\collaboration{PHENIX Collaboration}  \noaffiliation

\date{\today}


\begin{abstract}

Recently, the PHENIX Collaboration has published second- and 
third-harmonic Fourier coefficients $v_2$ and $v_3$ for midrapidity 
($|\eta|<0.35$) charged hadrons in 0\%--5\% central $p$$+$Au, $d$$+$Au, 
and $^3$He$+$Au collisions at $\sqrt{s_{_{NN}}}=200$ GeV utilizing three 
sets of two-particle correlations for two detector combinations with 
different pseudorapidity acceptance [Phys. Rev. C {\bf 105}, 024901 
(2022)]. This paper extends these measurements of $v_2$ to all 
centralities in $p$$+$Au, $d$$+$Au, and $^3$He$+$Au collisions, as well 
as $p$$+$$p$ collisions, as a function of transverse momentum ($p_T$) 
and event multiplicity. The kinematic dependence of $v_2$ is quantified 
as the ratio $R$ of $v_2$ between the two detector combinations as a 
function of event multiplicity for $0.5$$<$$p_T$$<$$1$ and 
$2$$<$$p_T$$<$$2.5$ GeV/$c$. A multiphase-transport (AMPT) model 
can reproduce the observed $v_2$ in most-central to midcentral $d$$+$Au 
and $^3$He$+$Au collisions.  However, the AMPT model 
systematically overestimates the measurements in $p$$+$$p$, $p$$+$Au, 
and peripheral $d$$+$Au and $^3$He$+$Au collisions, indicating a higher 
nonflow contribution in AMPT than in the experimental data.  The 
AMPT model fails to describe the observed $R$ for 
$0.5$$<$$p_T$$<$$1$ GeV/$c$, but there is qualitative agreement with the 
measurements for $2$$<$$p_T$$<$$2.5$ GeV/$c$.

\end{abstract}

\maketitle

\section{Introduction}

Observations of azimuthal anisotropy in the emission of produced 
particles in high-energy heavy-ion collisions at the Relativistic Heavy 
Ion Collider (RHIC) are considered to be strong evidence of the 
formation of the quark-gluon plasma (QGP)~\cite{Adcox:2004mh, 
Arsene:2004fa, Back:2004je,Adams:2005dq}. The measured anisotropy at 
RHIC and the Large Hadron Collider, quantified via Fourier 
coefficients \vn of the final-state particle yield relative to the 
participant plane, is successfully reproduced by viscous hydrodynamic 
calculations~\cite{Romatschke:2009im, Heinz:2013th}. These theoretical 
analyses of the experimental \vn data suggest that the collision 
geometry is translated into the final state momentum space via the 
hydrodynamic expansion of the QGP.

Heavy-ion experiments have also studied cold-nuclear-matter effects as 
potential backgrounds for QGP measurements, utilizing small collision 
systems, consisting of a light nucleus colliding with a heavy nucleus, 
where QGP formation had not been expected due to the small system size 
and low multiplicity. However, azimuthal anisotropy similar to that 
found in large collision systems has also been observed in 
high-multiplicity \ppb collisions at \sqsn~=~5.02~TeV at the Large 
Hadron Collider~\cite{ATLAS:2012cix,ALICE:2012eyl,CMS:2012qk} and in 
high-multiplicity \dau collisions at \sqsn~=~200~GeV at 
RHIC~\cite{PHENIX:2013ktj}. These surprising measurements raised the 
question of whether the \vn originates from the hydrodynamic expansion 
of the initial collision geometry in such small collision systems as 
well.

To address this question, it was proposed to experimentally examine the 
initial geometry dependence of the medium expansion, empirically known 
to hold in heavy-ion collisions, using the second- and third-harmonic 
azimuthal anisotropies \vtwo and \vthree~\cite{Nagle:2013lja}. For this 
purpose, from 2014 to 2016, RHIC delivered \pau, \dau, and \heau 
collisions at \sqsntwo. The series of \vn measurements with these data 
sets by the PHENIX Collaboration~\cite{PHENIX:2016cfs, 
Aidala:2017pup,PHENIX:2015idk,PHENIX:2014fnc}, culminating in the 
complete set of results published in Nature 
Physics~\cite{PHENIX:2018lia}, show that \vtwo and \vthree follow the 
pattern of the second- and third-harmonic initial eccentricities 
$\varepsilon_2$ and $\varepsilon_3$ estimated using the Monte 
Carlo-Glauber model. This observed relationship between initial geometry 
and final state correlations serves as evidence for QGP formation in 
small collision systems. The STAR Collaboration reported that 
$\vtwo/\varepsilon_2$, as a function of charged particle multiplicity to 
the minus-one-third power $\left<N_{ch}\right>^{-1/3}$, forms a common 
curve among high-multiplicity small- and large-system 
collisions~\cite{STAR:2019zaf}, which also implies the same underlying 
physics processes in such collision systems.

Additional hydrodynamic predictions with MC-Glauber initial 
conditions~\cite{Shen:2016zpp} also successfully reproduced the observed 
data, which corroborates formation of the QGP in small collision 
systems. Contrariwise, calculations based solely on initial-state 
correlations in the color-glass-condensate effective-field-theory 
formalism~\cite{Mace:2018vwq,Mace:2018yvl} are ruled out by the 
experimental data.

Furthermore, some hydrodynamic calculations incorporate the effect of 
prehydrodynamization parton dynamics with the 
weak~\cite{Schenke:2019pmk} and strong~\cite{Habich:2014jna} coupling 
limits. Both calculations are in quantitative agreement with the 
experimental data. However, the size of the prehydrodynamization 
dynamics cannot be determined with the current experimental and 
theoretical uncertainties. A systematic study of the collision-system 
and energy dependences in the hydrodynamic 
calculations~\cite{Romatschke:2015gxa} indicates the contribution of the 
prehydrodynamization dynamics becomes more pronounced in smaller 
collisions and at lower energies, where the QGP medium has a shorter 
lifetime. Extending experimental measurements to even smaller systems 
than high-multiplicity \pau, \dau, and \heau collisions can provide 
additional insights into the prehydrodynamization dynamics.

More recently, the PHENIX Collaboration has reported \vtwo and \vthree 
in 0\%--5\% central \pau, \dau, and \heau collisions at \sqsn~=~200~GeV 
obtained with three sets of two-particle correlations (2PC) for two 
detector combinations with different pseudorapidity 
acceptance~\cite{PHENIX:2021bxz}. One set of those measurements used the 
same detectors, i.e.~two detectors at backward rapidity (the Au-going 
direction) and one at midrapidity, and found good agreement between the 
3$\times$2PC method results and the event plane method results reported 
in Ref.~\cite{PHENIX:2018lia}. Another set of those measurements 
included a detector located at forward rapidity (\pdhe-going 
direction), which results in significantly larger \vtwo values and 
imaginary \vthree in \pau and \dau collisions. A careful 
analysis~\cite{Nagle:2021rep} of these experimental measurements 
suggests substantial nonflow contributions at forward rapidity because 
of both low multiplicity and possible longitudinal decorrelation 
effects. Estimating the multiplicity dependence of these effects would 
also be of interest to understand flow patterns in small systems.

In this article, our earlier \vtwo measurements~\cite{PHENIX:2021bxz} are 
extended from most-central to peripheral \pau, \dau, and \heau collisions, 
as well as \pp collisions, as a function of transverse momentum (\pt) and 
event multiplicity.  These measurements provide experimental data with 
different fractional contributions of prehydrodynamization, nonflow, and 
decorrelation effects. We also compare these measurements with a multiphase 
transport (AMPT) model~\cite{Lin:2004en} calculations, and the implications 
for nonflow and event-plane decorrelation effects in the kinematic selection 
dependence of \vtwo are discussed.

\section{Analysis Methodology}

This section details the detector subsystems of the PHENIX 
experiment, the analysis method employed, and the assessment of 
systematic uncertainties in this analysis.

\subsection{PHENIX Detectors}
\label{subsec:experiment}

The east and west central arms (CNT)~\cite{Adcox:2003zm} reconstruct 
charged particle tracks using the drift chambers and pad-chamber layers. 
Each arm covers a pseudorapidity range of $|\eta|<0.35$ with an 
azimuthal ($\phi$) coverage of $\pi/2$. The drift chambers determine 
the track momentum and the pad chambers reject background tracks by 
requiring that the track hits be within two standard deviations of their 
associated projections.  In this analysis, CNT tracks below 
$\pt=4$~GeV/$c$ are used to avoid background tracks from conversion 
electrons at high \pt.

The forward-silicon-vertex (FVTX) detectors~\cite{Aidala:2013vna} are 
installed in both the negative-rapidity south-side region (Au-going 
direction) and the positive-rapidity north-side region (\pdhe-going 
direction), covering $1<|\eta|<3$ with full $2\pi$ azimuthal acceptance. 
Both the south-side FVTX (FVTXS) and north-side FVTX (FVTXN) are used in 
this analysis. Charged particles within the acceptance of $1.2<|\eta|<2.2$ 
and transverse momentum of $\pt>0.3$~GeV/$c$ are reconstructed using the 
FVTX. The FVTX does not provide momentum information for tracks because of 
the orientation of the FVTX strips relative to the magnetic field. The FVTX 
also provides the distance of closest approach to the primary collision 
vertex in the transverse direction to the beam axis (${\rm DCA}_R$) with a 
resolution of 1.2 cm at \pt~=~0.5 GeV/$c$. Tracks with $|{\rm DCA}_R|<2$~cm 
are used in this analysis to reject background tracks.

Two beam-beam counters (BBC)~\cite{Allen:2003zt} are arrayed around the 
beam pipe at $\pm$144 cm from the nominal beam interaction point in both 
the south-side and north-side regions, covering the pseudorapidity range 
of $3.1<|\eta|<3.9$ with full $2\pi$ azimuthal acceptance. Each BBC 
comprises 64 \v{C}erenkov radiators equipped with a photomultiplier tube 
(PMT) and measures the total charge deposited in its acceptance, which 
is proportional to the number of particles.

The BBC triggers on minimum-bias (MB) \pp, \pau, \dau, and \heau 
collisions by requiring at least one hit on each side. The MB trigger 
efficiency is 55$\pm$5\%, 84$\pm$3\%, 88$\pm$4\%, and 88$\pm$4\% for 
inelastic \pp, \pau, \dau, and \heau collisions, respectively. Triggered 
events are further required to have an online $z$-vertex within 
$|z|<10$~cm in this analysis. The collision centralities in \pau, \dau, 
and \heau collisions are determined using the total charge in the 
south-side BBC (BBCS), as described in Ref.~\cite{Adare:2013nff}. The 
high-multiplicity trigger additionally required more than $35$, $40$, 
$49$ hit tubes in the BBCS for \pau, \dau, and \heau collisions, 
respectively. In Ref~\cite{PHENIX:2018lia}, the high-multiplicity 
trigger is used to improve the statistics of the 0\%--5\% centrality 
selection. In the present analysis, for more peripheral 
collisions only the MB trigger is used.

The instantaneous luminosities delivered by RHIC for \pp, \pau, \dau, and 
\heau collisions at \sqsn~=~200 GeV during 2014, 2015, and 2016 were high 
enough to record multiple collisions (i.e.~pileup). Typically multiple 
collisions occur at different positions along the beam direction, which is 
reflected as broader or secondary peaks in the timing distribution of hits 
in the BBCS. In each event, this shape is quantified as the fraction $f$ of 
the BBCS hits that have times within a 0.5 ns window from the most probable 
value of the measured timing distribution, as was done in 
Ref.~\cite{Aidala:2017pup}. Pileup events are rejected by requiring $f>0.9$.

\subsection{The 3$\times$2PC method}

In this analysis, the two-particle correlation method is employed. Because 
of the asymmetry in both the multiplicity and \vn as a function of 
pseudorapidity~\cite{PHENIX:2018hho}, two-particle 
azimuthal correlations are constructed with three different sets of pairs.
This method was developed in Ref.~\cite{PHENIX:2021bxz} and is called the 
3$\times$2PC method.

The 2PC function $C(\Delta\phi)$ is defined as
\begin{align}
C(\Delta\phi) &= \frac{S(\Delta\phi)}{M(\Delta\phi)}\frac{\int_0^{2\pi}d\Delta\phi M(\Delta\phi)}{\int_0^{2\pi}d\Delta\phi S(\Delta\phi)}, \\
S(\Delta\phi) &= \frac{dN_{\rm same}(\Delta\phi)\times w}{d\Delta\phi}, \\
M(\Delta\phi) &= \frac{dN_{\rm mixed}(\Delta\phi)\times w}{d\Delta\phi},
\end{align}
where $\Delta\phi$ is the difference in the azimuthal angles between two 
particles, $S(\Delta\phi)$ is the foreground distribution constructed 
from track pairs in the same event $N_{\rm same}$, and $M(\Delta\phi)$ is 
the mixed event distribution constructed from track pairs from different 
events $N_{\rm mixed}$ in the same centrality and collision vertex class. 
The weight $w$ is 1 when correlating with tracks and the charge in the 
PMT when correlating with BBC PMTs.

We fit the correlation functions with a Fourier series up to the fourth 
harmonic:
 \begin{equation}
 F(\Delta\phi) = 1 + \sum_{n=1}^4 2 c_n \cos{n\Delta\phi},
 \end{equation}
where $c_n=\left<\cos{n\Delta\phi}\right>$ is the $n$-th harmonic 
Fourier component and $n$ is the harmonic number. Under the 
flow-factorization assumption, the obtained $c_n$ can be related to \vn as
\begin{align}
c_n^{AB} &= \langle  v_n^Av_n^B\rangle, \\
c_n^{AC} &= \langle  v_n^Av_n^C\rangle, \\
c_n^{BC} &= \langle  v_n^Bv_n^C\rangle,
\end{align}

\noindent where $A$, $B$, and $C$ stand for sub events used to measure 
correlation functions. Finally, \vn is obtained as
\begin{equation}
v_n^C\{{\rm 3}\times{\rm 2PC}\}(\pt^C) = \sqrt{\frac{c_n^{AC}(\pt^C)\times c_n^{BC}(\pt^C)}{c_n^{AB}}},
\label{eq:2pc_abc}
\end{equation}
letting the sub-event $C$ be CNT for the midrapidity \vn measurements 
presented in this manuscript. Here we assume that detector effects in the sub events
$A$ and $B$ are canceled out between the numerator and denominator inside the square root
of Eq.~(\ref{eq:2pc_abc}).


\begin{figure}[tbh]
\includegraphics[width=0.95\linewidth]{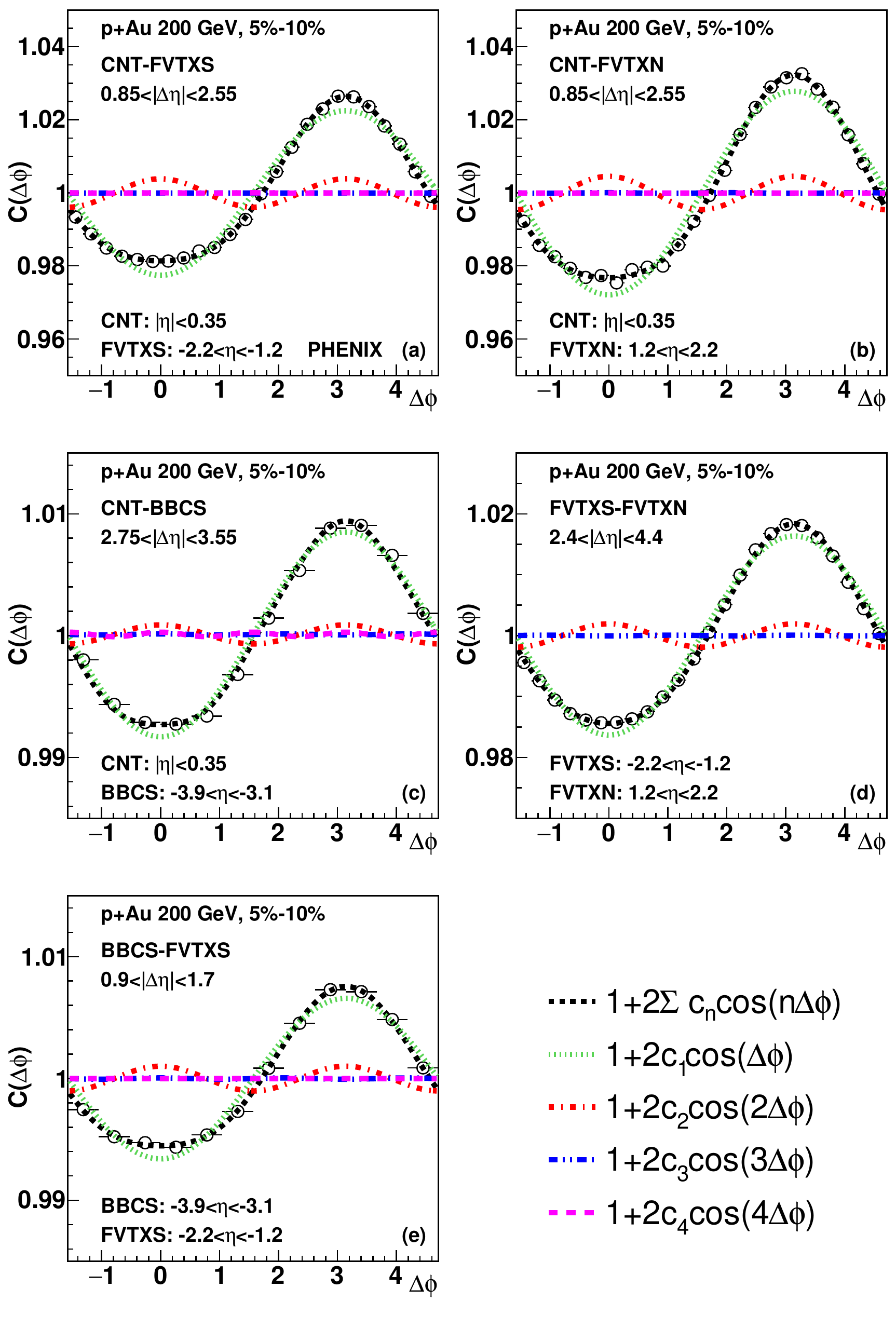}
\caption{
Correlation functions $C(\Delta\phi)$ in 5\%--10\% centrality \pau 
collisions at \sqsntwo measured using (a) CNT-FVTXS, (b) CNT-FVTXN, (c) 
CNT-BBCS, (d) FVTXS-FVTXN, and (e) BBCS-FVTXS detector combinations. The 
short-dashed [black] curve shows the Fourier fit to correlation functions. 
The dotted [green], dash-dotted [red], dashed-double-dotted [blue], and 
long-dashed [magenta] curves indicate $c_1$, $c_2$, $c_3$, and $c_4$ components, 
respectively.
}
 \label{fig:pau_cor_5-10}
 \end{figure}

\begin{figure*}[tbh] 
\begin{minipage}[tbh]{0.48\linewidth}
\includegraphics[width=0.99\linewidth]{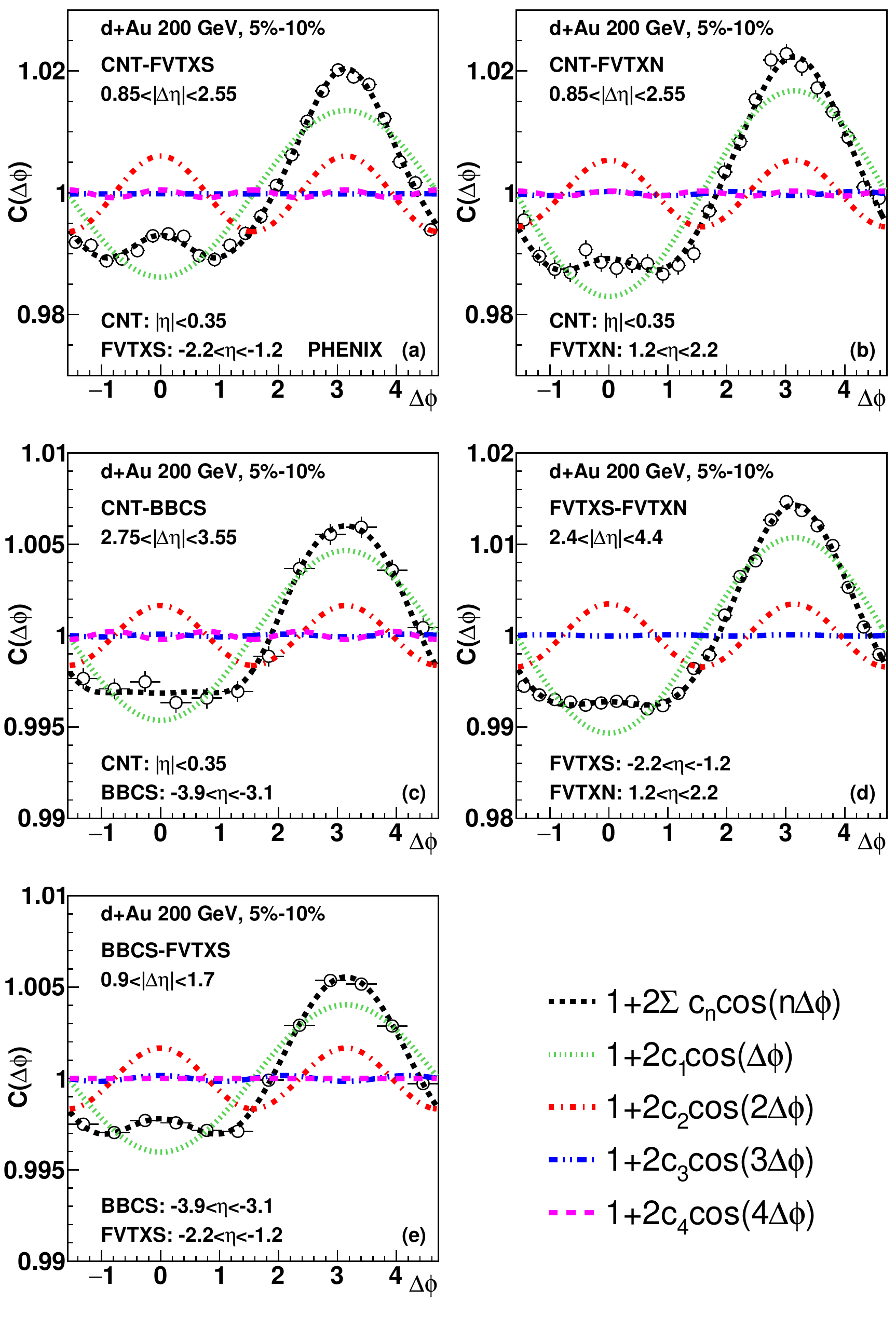}
\caption{
Correlation functions $C(\Delta\phi)$ in 5\%--10\% centrality \dau 
collisions at \sqsntwo measured using (a) CNT-FVTXS, (b) CNT-FVTXN, (c) 
CNT-BBCS, (d) FVTXS-FVTXN, and (e) BBCS-FVTXS detector combinations. The 
short-dashed [black] curve shows the Fourier fit to correlation functions. 
The dotted [green], dash-dotted [red], dashed-double-dotted [blue], and 
long-dashed [magenta] curves indicate $c_1$, $c_2$, $c_3$, and $c_4$ components, 
respectively.
}
\label{fig:dau_cor_5-10}
\end{minipage}
\hspace{0.2cm}
\begin{minipage}[tbh]{0.48\linewidth}
\includegraphics[width=0.99\linewidth]{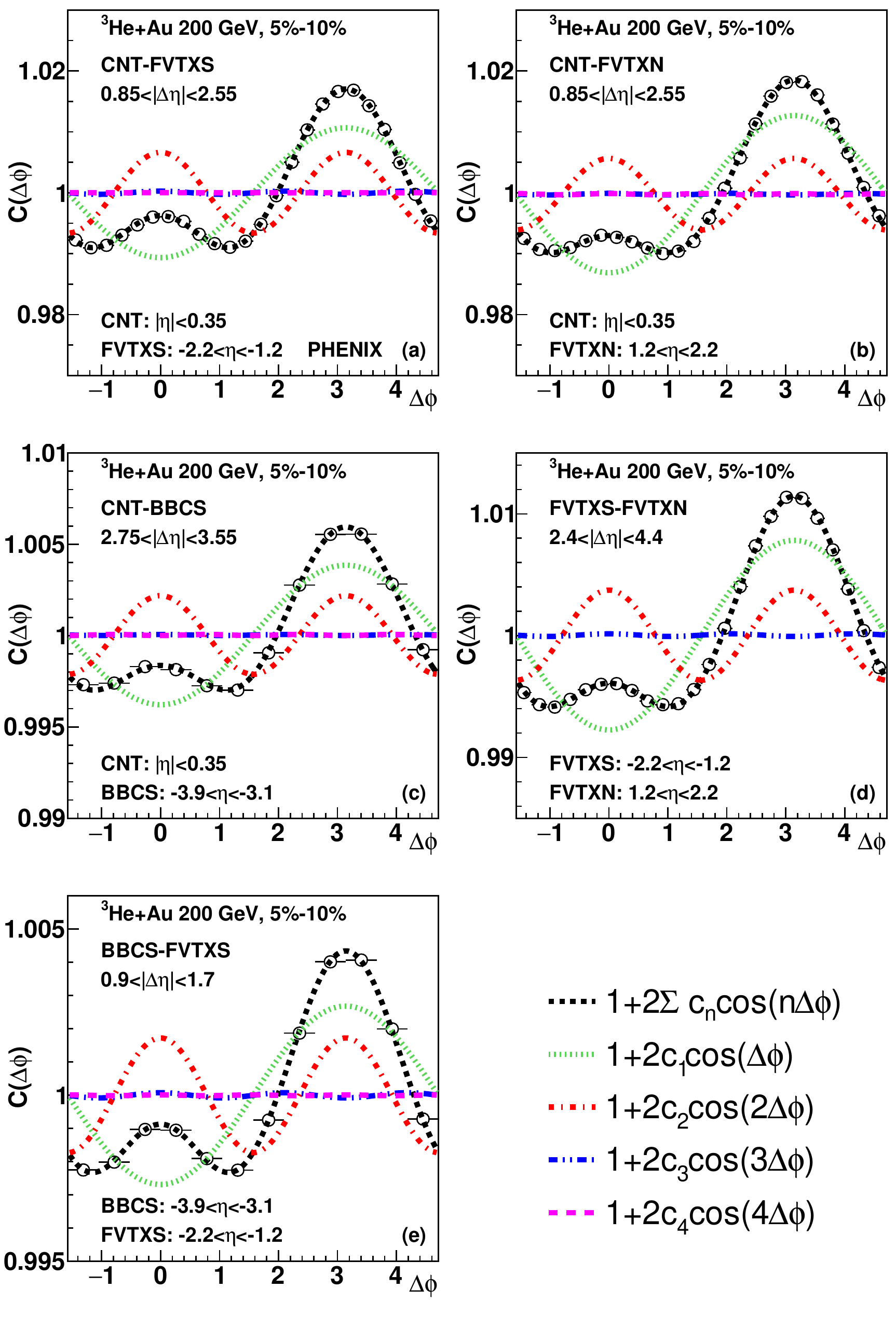}
\caption{
Correlation functions $C(\Delta\phi)$ in 5\%--10\% centrality \heau 
collisions at \sqsntwo measured using (a) CNT-FVTXS, (b) CNT-FVTXN, (c) 
CNT-BBCS, (d) FVTXS-FVTXN, and (e) BBCS-FVTXS detector combinations. The 
short-dashed [black] curve shows the Fourier fit to correlation functions. 
The dotted [green], dash-dotted [red], dashed-double-dotted [blue], and 
long-dashed [magenta] curves indicate $c_1$, $c_2$, $c_3$, and $c_4$ components, 
respectively.
}
\label{fig:heau_cor_5-10}
\end{minipage}
\end{figure*}

Figures~\ref{fig:pau_cor_5-10},~\ref{fig:dau_cor_5-10}, 
and~\ref{fig:heau_cor_5-10} show $C(\Delta\phi)$ and the Fourier fits to 
$C(\Delta\phi)$ in 5\%--10\% central \pau, \dau, and \heau collisions at 
\sqsn~=~200 GeV, respectively. In each panel of 
Figs.~\ref{fig:pau_cor_5-10},~\ref{fig:dau_cor_5-10}, 
and~\ref{fig:heau_cor_5-10}, correlations are measured between
\vspace{-\topsep}
\begin{itemize}
  \setlength{\parskip}{0pt}
  \setlength{\itemsep}{0pt plus 1pt}
\item[(a)] CNT tracks and FVTXS tracks,
\item[(b)] CNT tracks and FVTXN tracks,
\item[(c)] CNT tracks and BBCS tubes,
\item[(d)] FVTXS and FVTXN tracks, and
\item[(e)] BBCS tubes and FVTXS tracks,
\end{itemize}

\noindent where CNT tracks are required to be $0.2<\pt<4$~GeV/$c$. The rapidity 
coverage of these detectors and rapidity gaps between the detector pairs 
used for the correlation functions are specified in each panel. See also 
Ref.~\cite{PHENIX:2021bxz} for the correlation functions in MB \pp and 
0\%--5\% central \pau, \dau, and \heau collisions.

Notably, a nonzero value of the second-harmonic coefficient $c_2$ is 
observed also in noncentral collisions for these correlation functions. 
Thus \vtwo can be measured in noncentral collisions with the 
3$\times$2PC method using the BBCS-FVTXS-CNT and FVTXS-CNT-FVTXN 
detector combinations as done for 0\%--5\% collisions in 
Ref.~\cite{PHENIX:2021bxz}. The former combination BBCS-FVTXS-CNT is 
denoted as ``BB'' as it uses two detectors located at backward rapidity. 
Similarly, the latter combination FVTXS-CNT-FVTXN is called ``BF'' as it 
uses one detector at backward rapidity and another detector at forward 
rapidity.

\begin{figure*}[tbh]
\begin{minipage}{0.99\linewidth}
\includegraphics[width=0.99\linewidth]{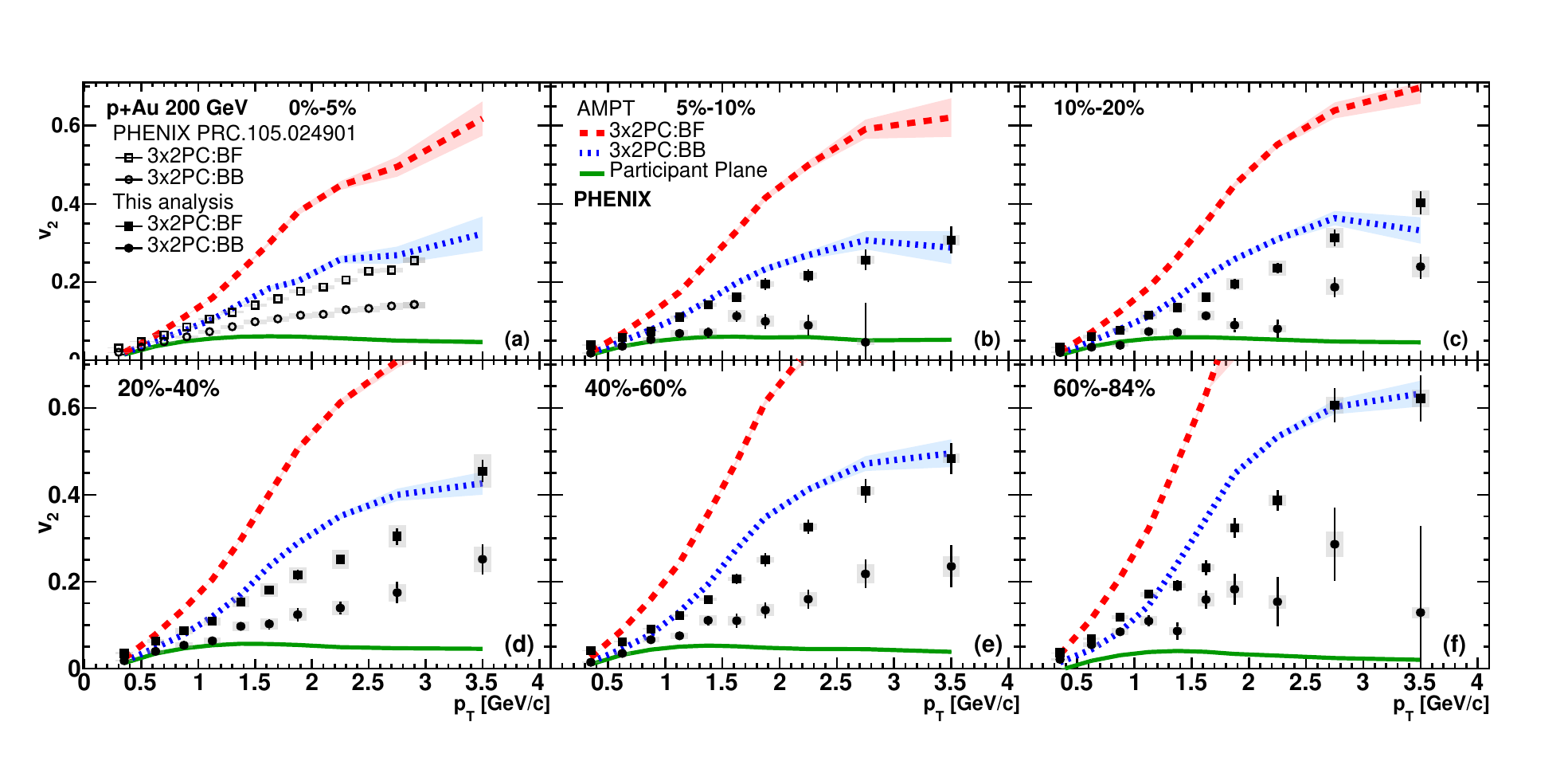}
\caption{
Second-harmonic azimuthal anisotropy $v_2\{\rm3\times2PC\}$ in (a) 
0\%--5\%~\protect\cite{PHENIX:2021bxz}, (b) 5\%--10\%, (c) 10\%--20\%, (d) 
20\%--40\%, (e) 40\%--60\%, and (f) 60\%--88\% centrality \pau collisions at 
\sqsntwo with the FVTXS-CNT-FVTXN (BF) and BBCS-FVTXS-CNT (BB) detector 
combinations as a function of \pt. The solid [black] squares are shifted for 
visibility. The bands around the [black] squares and [black] circles show 
the systematic uncertainties. The bands around the dashed [red] and dotted 
[blue] curves show statistical uncertainties in the AMPT calculations with 
the 3$\times$2PC method. The solid [green] curves show $v_2$ in AMPT using 
the parton participant plane.
}
\label{fig:pau_v2_pt_centrality}
\end{minipage}
\begin{minipage}{0.99\linewidth}
\includegraphics[width=0.99\linewidth]{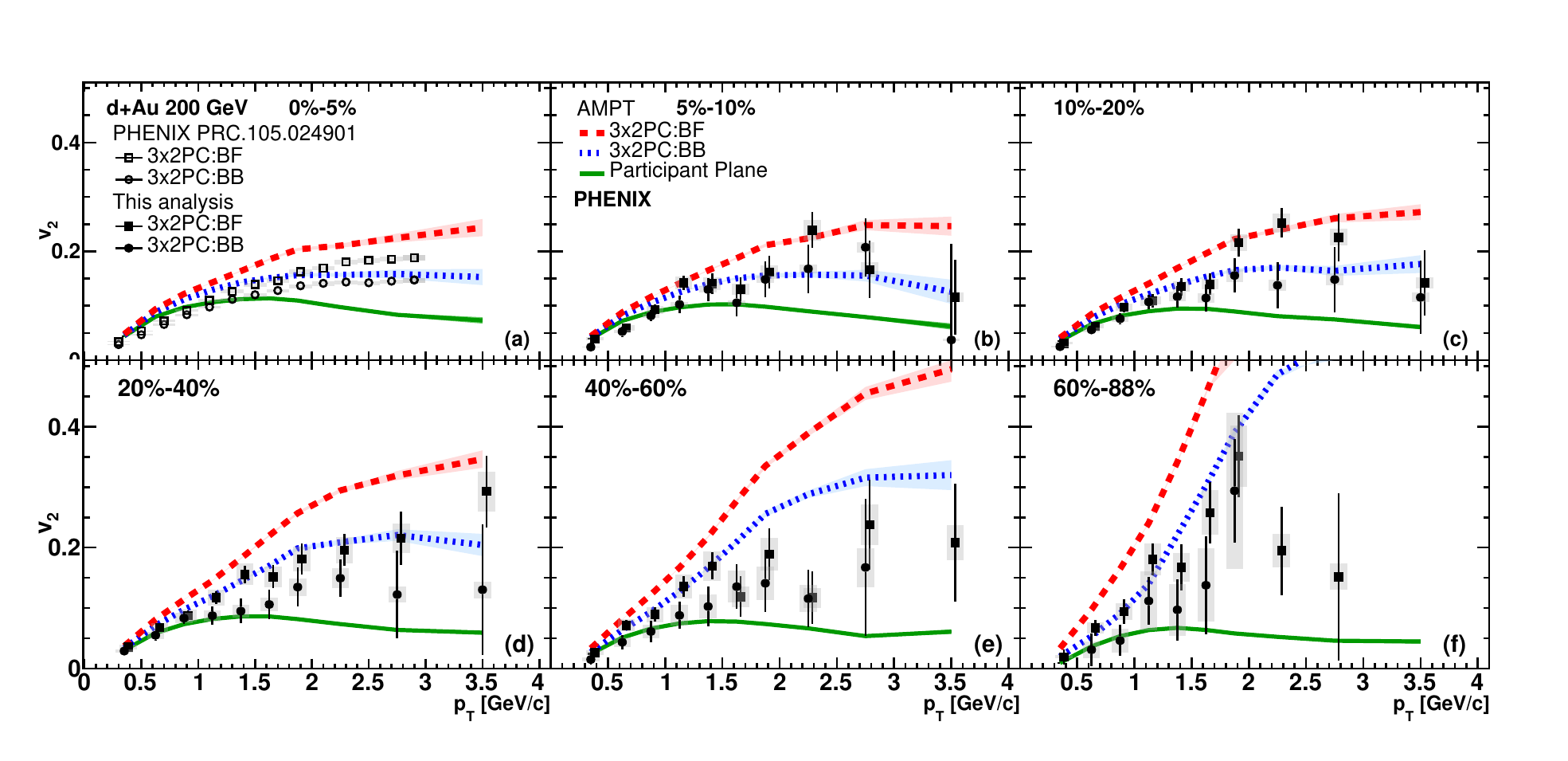}
\caption{
Second-harmonic azimuthal anisotropy $v_2\{\rm3\times2PC\}$ in (a) 
0\%--5\%~\protect\cite{PHENIX:2021bxz}, (b) 5\%--10\%, (c) 10\%--20\%, (d) 
20\%--40\%, (e) 40\%--60\%, and (f) 60\%--88\% centrality \dau collisions at 
\sqsntwo with the FVTXS-CNT-FVTXN (BF) and BBCS-FVTXS-CNT (BB) detector 
combinations as a function of \pt. The solid [black] squares are shifted for 
visibility. The bands around the [black] squares and [black] circles show 
the systematic uncertainties. The bands around the dashed [red] and dotted 
[blue] curves show statistical uncertainties in the AMPT calculations with 
the 3$\times$2PC method. The solid [green] curves show $v_2$ in AMPT using 
the parton participant plane.
}
\label{fig:dau_v2_pt_centrality}
\end{minipage}
\end{figure*}

\begin{figure*}[tbh]
\includegraphics[width=0.99\linewidth]{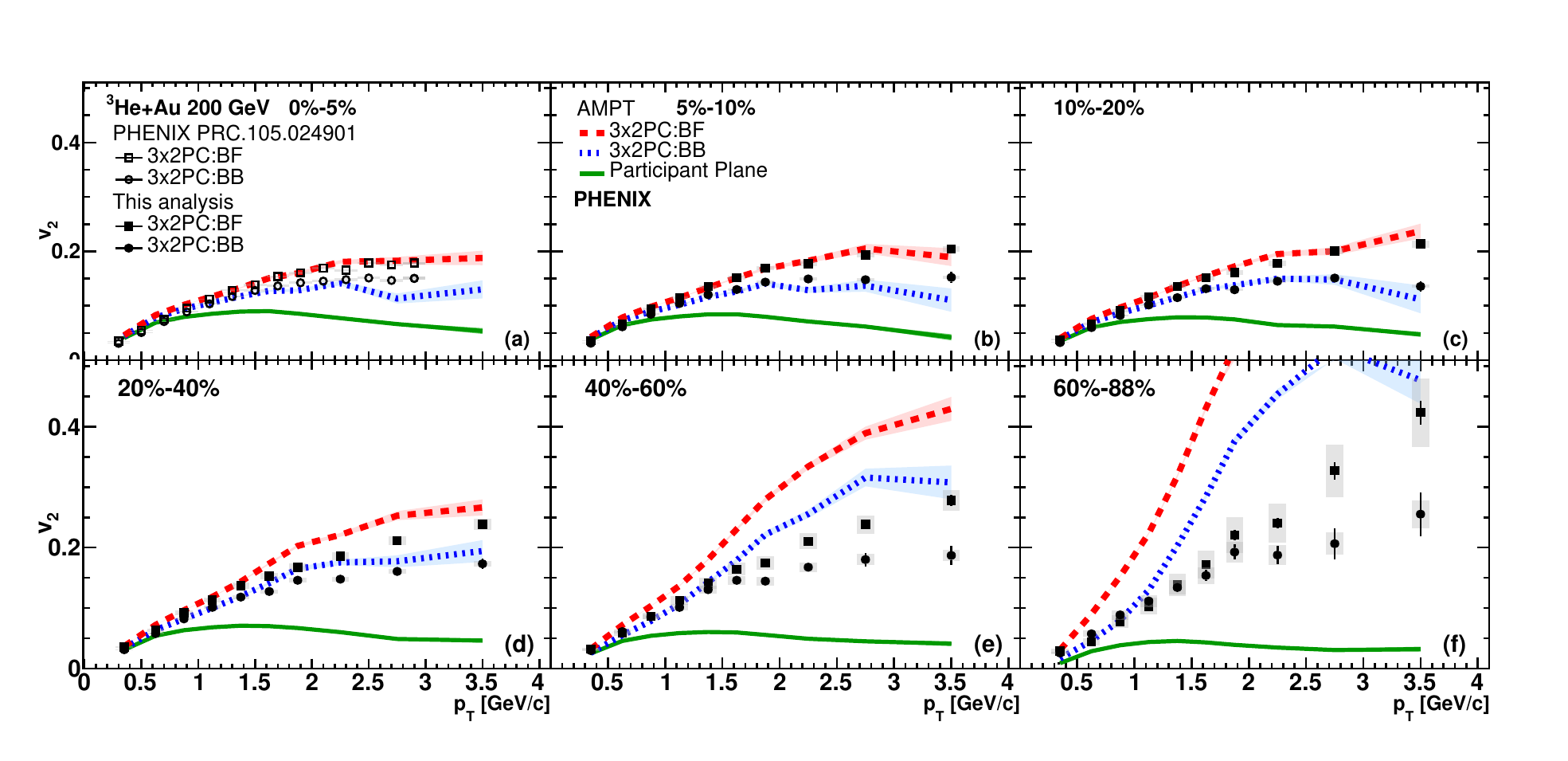}
\caption{
Second-harmonic azimuthal anisotropy $v_2\{\rm3\times2PC\}$ in (a) 
0\%--5\%~\protect\cite{PHENIX:2021bxz}, (b) 5\%--10\%, (c) 10\%--20\%, (d) 
20\%--40\%, (e) 40\%--60\%, and (f) 60\%--88\% centrality \heau collisions at 
\sqsntwo with the FVTXS-CNT-FVTXN (BF) and BBCS-FVTXS-CNT (BB) detector 
combinations as a function of \pt. The solid [black] squares are shifted for 
visibility. The bands around the [black] squares and [black] circles show 
the systematic uncertainties. The bands around the dashed [red] and dotted 
[blue] curves show statistical uncertainties in the AMPT calculations with 
the 3$\times$2PC method. The solid [green] curves show $v_2$ in AMPT using 
the parton participant plane.
}
\label{fig:heau_v2_pt_centrality}
\end{figure*}
\subsection{Systematic Uncertainty}

In this analysis, systematic uncertainties on the measured \vtwo are 
considered for the CNT arm selection, pad-chamber matching width, FVTX track 
${\rm DCA}_R$, and pileup rejection using the timing information of hit tubes in 
the BBCS. The central \vtwo values are calculated using both the east and 
west CNT arms, pad-chamber matching width of 2$\sigma$, $|{\rm DCA}_R|<2$~cm, and 
BBC timing fraction $f>0.9$. The systematic uncertainty associated with CNT 
arm selection is obtained from the difference between \vtwo in the east and 
west CNT arms. The systematic uncertainty associated with the pad-chamber 
matching is estimated by varying the matching width from 1.5$\sigma$ to 
2.5$\sigma$. The systematic uncertainty associated with the FVTX ${\rm DCA}_R$ cut 
is estimated by varying the ${\rm DCA}_R$ cut from 1.5~cm~to~2.5~cm. Finally, the 
systematic uncertainty associated with pileup rejection is estimated by 
varying the BBC-timing-fraction cut from $f>0.85$ to $f>0.95$. Given the 
limited statistical precision at high-\pt, the systematic uncertainty is 
determined for $\pt<3$~GeV/$c$ and is applied to the entire \pt region.

The CNT arm selection is the largest source of systematic uncertainty 
and has an effect of up to 12\% depending on collision system and 
centrality. The pad-chamber matching window and BBC-timing-fraction cuts 
have effects on the order of a few percent. The FVTX ${\rm DCA}_R$ cuts have an 
effect of less than one percent in most cases. Each systematic 
uncertainty is added in quadrature to obtain the total systematic 
uncertainty.

\section{Results}

The experimental \vtwo for midrapidity charged particles in \pp, 
\pau, \dau, and \heau collisions at \sqsntwo is presented as a function of 
\pt, centrality, and event multiplicity.  Then, the experimental results 
are compared to AMPT-model simulations and physics implications are 
discussed.  Noting that previous flow extractions were restricted to 0\%--5\% 
central \pau, \dau, and \heau collisions, estimates of 
nonflow contributions indicated flow dominance. In the present analysis, 
pushing to lower multiplicities, including \pp collisions, it is expected 
that nonflow will have a larger role and become dominant, for example in \pp 
collisions. Thus, extraction of the second Fourier coefficient as $v_2$ 
should not necessarily be interpreted as flow, but rather an interplay of 
different effects.

\subsection{\pt Dependence}

Shown in Figs.~\ref{fig:pau_v2_pt_centrality}, 
\ref{fig:dau_v2_pt_centrality}, and~\ref{fig:heau_v2_pt_centrality} is 
\vtwo with the 3$\times$2PC method as a function of \pt in different 
centrality selections for \pau, \dau, and \heau collisions at \sqsntwo, 
respectively. The results in the 0\%--5\% most-central collisions are 
from Ref.~\cite{PHENIX:2021bxz}. Notably, nonzero \vtwo is observed 
over the entire measured \pt range from most-central to most-peripheral 
collisions in these systems, with both the BB and BF detector 
combinations.

The kinematic dependence seen in 0\%--5\% central collisions, 
i.e.~larger $v_2\{\rm 3\times2PC\}$ with the BF combination ($v_2\{{\rm 
BF}\}$) than that with the BB combination ($v_2\{{\rm BB}\}$), is also 
observed in noncentral \pau and \heau collisions. This trend becomes 
visible above $\pt=0.5$~GeV/$c$ in \pau collisions and above 
$\pt=1.5$~GeV/$c$ in \heau collisions. These observations in noncentral 
\pau and \heau collisions confirm the interpretation of the kinematic 
dependence discussed in Ref.~\cite{PHENIX:2021bxz}: the smaller 
multiplicity in the FVTXN acceptance relative to that in the BBCS 
acceptance results in more nonflow which makes the observed \vtwo 
larger. The larger rapidity gap between FVTXS and FVTXN compared to that 
between BBCS and FVTXS also increases the event-plane decorrelation 
effects, which makes the denominator of Eq.~(\ref{eq:2pc_abc}) smaller. 
However, the factorization of the decorrelation effects between the 
numerator and denominator is under discussion~\cite{Nagle:2021rep} and 
thus the influence on \vtwo is inconclusive. In contrast, the relation 
of $v_2\{{\rm BF}\}=v_2\{{\rm BB}\}$ holds below $\pt<1.5$~GeV/$c$ in 
\heau collisions. Note that no kinematic dependence is observed in 
noncentral \dau collisions due to the limited statistical precision.

\begin{figure}[tbh]
\includegraphics[width=0.95\linewidth]{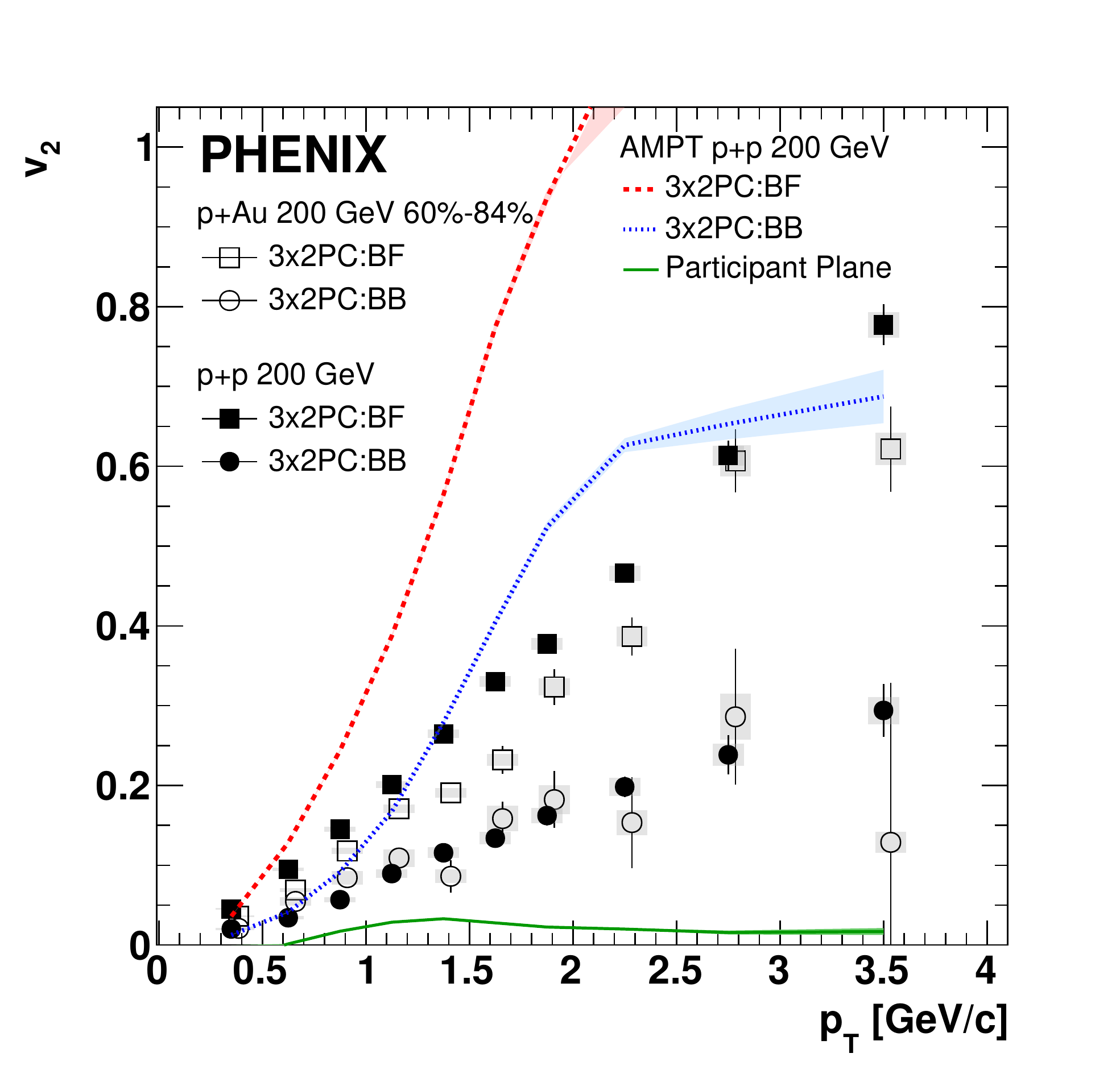}
\caption{
Second-harmonic azimuthal anisotropy \vtwo with the 3$\times$2PC method in 
(open symbols) 60\%--84\% central \pau collisions and (solid symbols) MB \pp 
collisions at \sqsntwo with the FVTXS-CNT-FVTXN (BF) and BBCS-FVTXS-CNT (BB) 
detector combinations as a function of \pt. The open [black] squares and 
[black] circles are shifted for visibility.  The solid bands around the 
[black] circles and [black] squares show experimental systematic 
uncertainties. The bands around the dashed [red] and dotted [blue] curves 
show statistical uncertainties in the AMPT calculations with the 
3$\times$2PC method in \pp collisions. The solid [green] curve shows $v_2$ in 
AMPT using the parton participant plane in \pp collisions.
}
\label{fig:pp_v2_pt}
\end{figure}

Measurement of \vtwo with the 3$\times$2PC method is further extended 
to MB \pp collisions as shown in Fig.~\ref{fig:pp_v2_pt}. Similar to the 
other collision systems, nonzero \vtwo is observed over the entire 
measured \pt range for both the BB and BF detector combinations. At 
$\pt=3.5$~GeV/$c$, the value of $v_2\{{\rm BB}\}$ remains at 0.3 while 
that of $v_2\{{\rm BF}\}$ soars to 0.8. The latter value larger than 0.5 
indicates that correlations from back-to-back jets are dominant in this 
kinematic range. The magnitude of \vtwo in \pp collisions is found to be 
similar to that of \vtwo in 60\%--84\% central \pau collisions.

\subsection{Multiplicity Dependence}

\begin{figure*}[tbh]
\begin{minipage}{0.67\linewidth}
\includegraphics[width=0.99\linewidth]{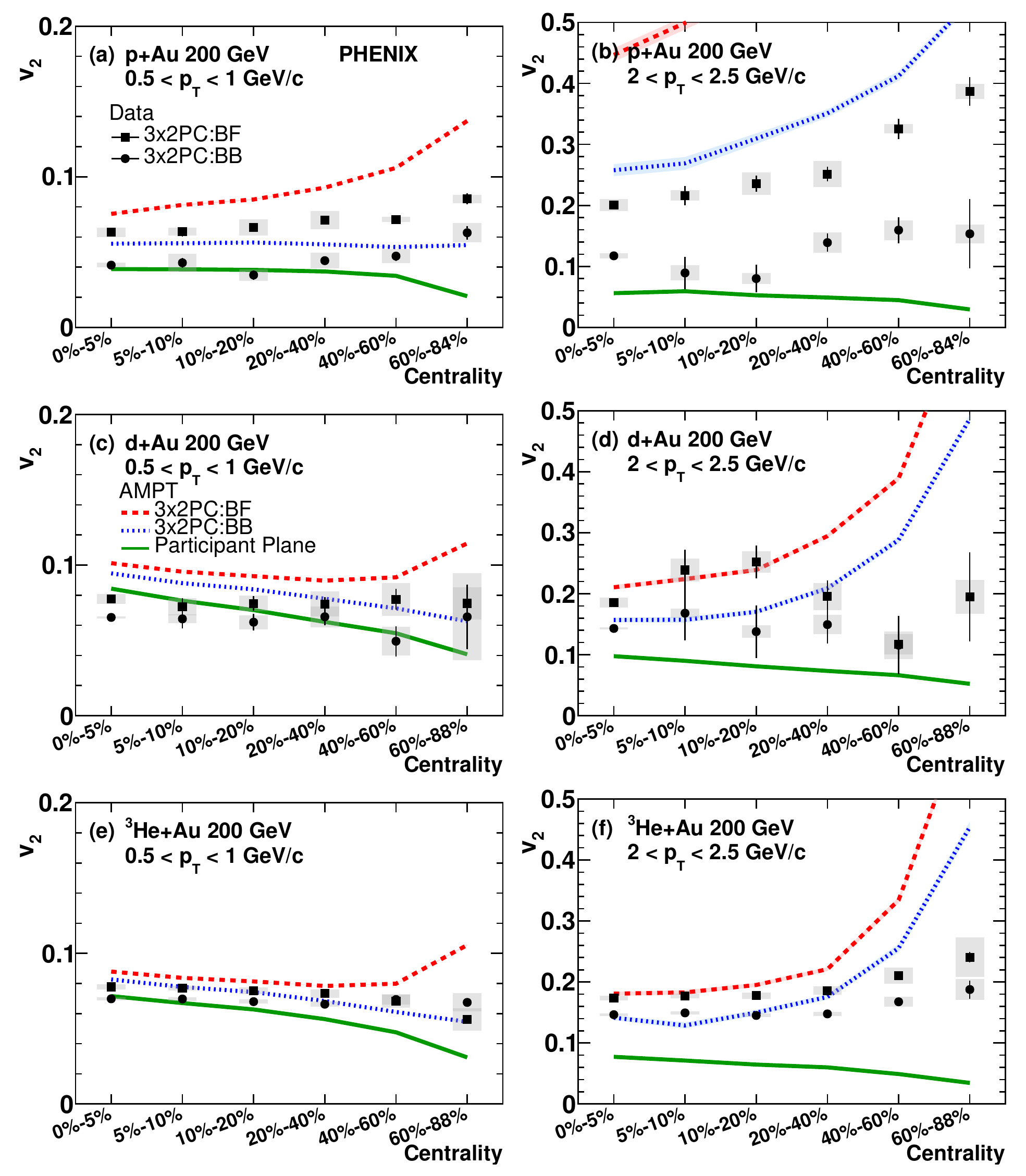}
\end{minipage}
\hspace{0.2cm}
\begin{minipage}{0.3\linewidth}
\caption{
Second-harmonic azimuthal anisotropy $v_2\{\rm3\times2PC\}$ as a function of 
centrality in (a,b) \pau, (c,d) \dau, and (e,f) \heau collisions at \sqsntwo 
with the FVTXS-CNT-FVTXN (BF) and BBCS-FVTXS-CNT (BB) detector combinations. 
The bands around the [black] circles and [black] squares show experimental 
systematic uncertainties. The bands around the dashed [red] and dotted 
[blue] curves show statistical uncertainties in the AMPT calculations with 
the 3$\times$2PC method. The solid [green] curves show $v_2$ in AMPT using 
the parton participant plane.
}
\label{fig:v2_centrality}
\end{minipage}
\end{figure*}

Figure~\ref{fig:v2_centrality} shows \vtwo with the 3$\times$2PC method 
in $0.5<\pt<1$~GeV/$c$ and $2<\pt<2.5$~GeV/$c$ as a function of 
centrality in \pau, \dau, and \heau collisions. In \dau and \heau 
collisions, \vtwo in $0.5<\pt<1$~GeV/$c$ is generally flat over the 
entire measured centrality range within uncertainties. Only \vtwo in 
\pau collisions shows an increasing trend towards peripheral collisions 
for both the BB and BF detector combinations. In $2<\pt<2.5$~GeV/$c$, 
\vtwo in \pau and \heau collisions show increasing trends towards 
peripheral collisions for both the BB and BF detector combinations. In 
\dau collisions, this trend is not observed because of the limited 
statistical precision.

\begin{figure*}[tbh]
\begin{minipage}[tbh]{0.67\linewidth}
\includegraphics[width=0.99\linewidth]{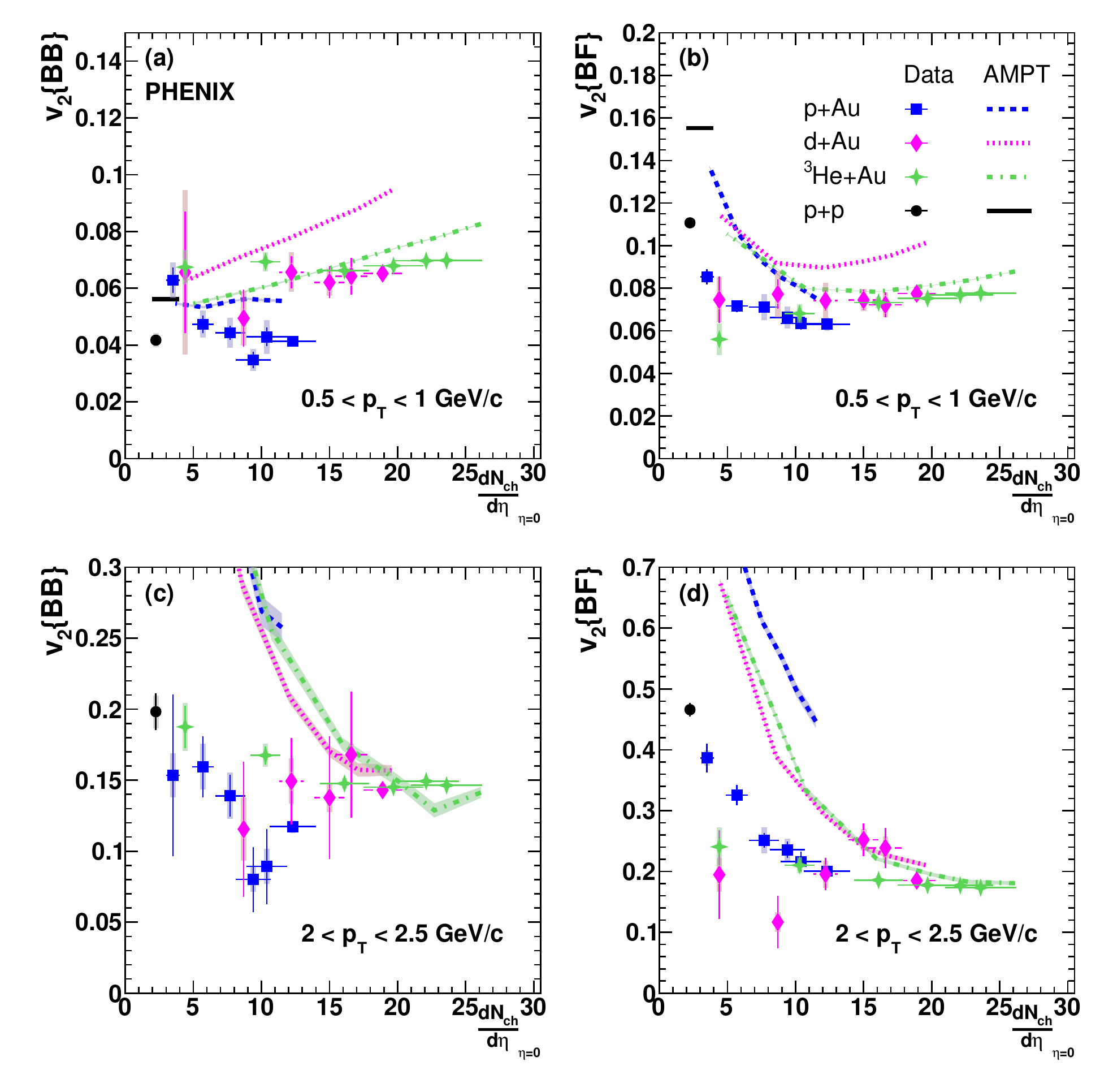}
\end{minipage}
\hspace{0.2cm}
\begin{minipage}[tbh]{0.3\linewidth}
\caption{
Second-harmonic azimuthal anisotropy $v_2\{\rm3\times2PC\}$ as a 
function of charged-particle multiplicity $\frac{dN_{ch}}{d\eta}$ at 
midrapidity in \pp, \pau, \dau, and \heau collisions at~\sqsn~=~200~GeV 
with (a,c) the BBCS-FVTXS-CNT (BB) and (b,d) FVTXS-CNT-FVTXN (BF) 
detector combinations. The bands around the data points show 
experimental systematic uncertainties and the bands around the curves 
show statistical uncertainties in the AMPT calculations. Note that 
AMPT results for \pp collisions in (c) and (d) are outside of the plot 
range due to their large values.
}
\label{fig:v2_vs_mult}
\end{minipage}
\end{figure*}

Figure~\ref{fig:v2_vs_mult} shows that a point-by-point comparison among the 
different collision systems can be made with the 3$\times$2PC method using 
both the BB and BF detector combinations by plotting \vtwo as a function of 
charged particle multiplicity $\frac{dN_{ch}}{d\eta}$ at midrapidity.  The 
values of $\frac{dN_{ch}}{d\eta}$ are obtained from 
Ref.~\cite{PHENIX:2018hho}. In $2<\pt<2.5$~GeV/$c$, $v_2\{{\rm BB}\}$ shows 
an increasing trend towards the low $\frac{dN_{ch}}{d\eta}$ side; the 
peripheral \pau data points smoothly connect to the \pp data point within 
uncertainties. This trend is more clearly seen in $v_2\{{\rm BF}\}$ for both 
$0.5<\pt<1$~GeV/$c$ and $2<\pt<2.5$~GeV/$c$. Above 
$\frac{dN_{ch}}{d\eta}=10$, these series of \vtwo measurements generally 
show flat trends. Unlike these trends, $v_2\{{\rm BB}\}$ in 
$0.5<\pt<1$~GeV/$c$ shows a flat shape over the entire measured 
$\frac{dN_{ch}}{d\eta}$ range within the current experimental uncertainties, 
which might indicate that the balance of nonflow effects between the 
numerator and denominator of Eq.~(\ref{eq:2pc_abc}) stays the same in this 
$\frac{dN_{ch}}{d\eta}$ range.

\begin{figure}[tbh]
\includegraphics[width=1.0\linewidth]{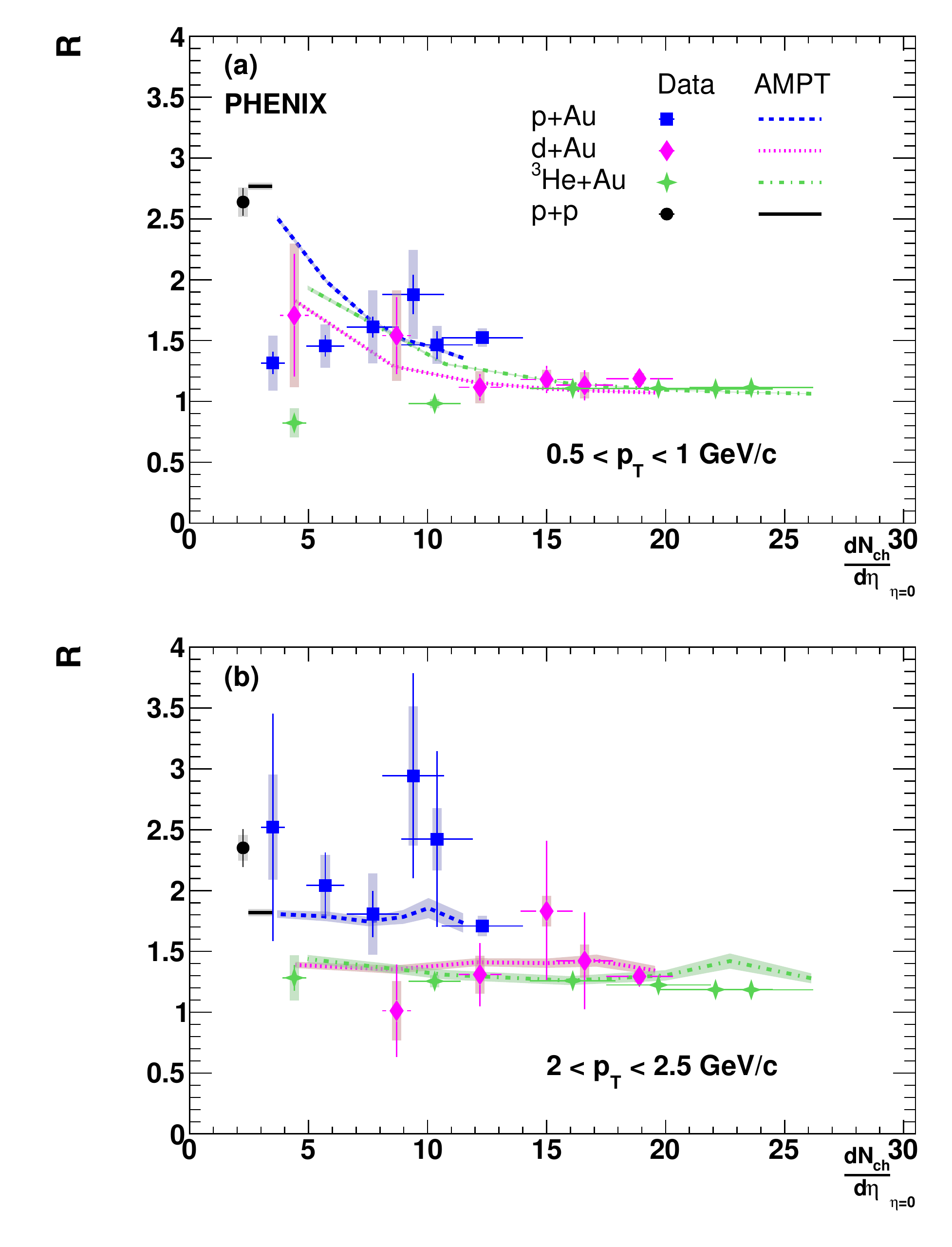}
\caption{
The ratio $R$ of $v_2\{{\rm BF}\}$ to $v_2\{{\rm BB}\}$ as a function of 
charged-particle multiplicity $\frac{dN_{ch}}{d\eta}$ at midrapidity in 
(squares) \pau, (diamonds) \dau, (crosses) \heau and (circles) \pp 
collisions at~\sqsn~=~200~GeV. The bands around the data points show 
experimental systematic uncertainties and the bands around the curves show 
statistical uncertainties in the AMPT calculations.
}
\label{fig:r_vs_mult}
\end{figure}

Finally, the kinematic dependence of \vtwo is quantified by the ratio $R$ 
of \vtwo in the BF detector combination to that in the BB combination. 
Figure~\ref{fig:r_vs_mult} shows $R$ as a function of charged-particle 
multiplicity $\frac{dN_{ch}}{d\eta}$ at midrapidity for 
$0.5<\pt<1$~GeV/$c$ and $2<\pt<2.5$~GeV/$c$. In $0.5<\pt<1$~GeV/$c$, $R$ 
in \dau and \heau collisions approaches unity as $\frac{dN_{ch}}{d\eta}$ 
increases, indicating weak kinematic dependence, i.e.~the restoration of 
flow factorization. Towards the low $\frac{dN_{ch}}{d\eta}$ side, $R$ in 
\heau collisions falls below unity, however $R$ in \pau and \dau 
collisions do not show clear trends due to the limited statistical and 
systematic precision. At the lowest $\frac{dN_{ch}}{d\eta}$, $R$ in \pp 
collisions shows the largest value among these collision systems. In 
$2<\pt<2.5$~GeV/$c$, the $R$ values in \dau and \heau collisions are 
consistent within uncertainties in the overlapping 
$\frac{dN_{ch}}{d\eta}$ region.  The measured $R$ is generally 
larger in \pau than in \dau and \heau collisions even in the overlapping 
$\frac{dN_{ch}}{d\eta}$ ranges. For the lowest values of 
$\frac{dN_{ch}}{d\eta}$, the values of $R$ in \pp and \pau collisions 
are consistent within uncertainties.  The different trends of $R$ between 
$0.5<\pt<1$~GeV/$c$ and $2<\pt<2.5$~GeV/$c$ likely indicate that the 
kinematic dependence is caused by different underlying mechanisms.

\subsection{Comparison With AMPT Model Simulations}

To further investigate the experimental \vtwo results, the AMPT model is 
employed with string melting turned on and the parton-parton interaction 
cross section set to 1.5~mb. We used the same AMPT parameter settings as 
those used in Ref.~\cite{Aidala:2017pup} for its \vtwo study in the \dau 
beam energy scan. In this AMPT model calculation, final-state particle \vtwo 
is calculated using the 3$\times$2PC method with the same \pt and rapidity 
range selections as the experimental measurements, as well as relative to 
the parton participant plane determined using initial partons. We use the 
parton participant plane \vtwo as a proxy of pure collective development of 
the collision system, which is likely to underestimate the true \vtwo value. 
The difference between \vtwo relative to the parton participant plane and 
that with the 3$\times$2PC method in AMPT model can provide some insight on 
the relative contributions from nonflow and event-plane decorrelation 
effects. Note that the experimental event trigger efficiency has not been 
applied to peripheral small systems and \pp collisions in this AMPT 
simulation and thus the full inelastic cross section was used in this study.

\subsubsection{\pt Dependence}

Figures~\ref{fig:pau_v2_pt_centrality},~\ref{fig:dau_v2_pt_centrality}, 
and~\ref{fig:heau_v2_pt_centrality} show comparisons of AMPT \vtwo 
with the experimental measurements as a function of \pt. The \vtwo 
calculated from AMPT with the 3$\times$2PC method generally describes 
the experimental \vtwo results from most-central to midcentral \dau and 
\heau collisions. However, it overshoots the data in all centralities for 
\pau collisions and in midcentral to peripheral centralities for \dau and \heau 
collisions, similar to what was previously reported in 
Ref~\cite{Aidala:2017pup} for peripheral \dau collisions, indicating much 
higher levels of nonflow in AMPT compared to the data. An explanation for 
this overestimate is that the HIJING model, used to describe 
hard-scattering processes in AMPT, is known to have a wider near-side 
jet correlation than in real \pp data~\cite{Lim:2019cys}. This mismatch of 
the jet kinematics leads to this overestimate. While \vtwo relative to the 
parton participant plane weakly depends on \pt, its difference from \vtwo 
with the 3$\times$2PC method increases with increasing \pt, indicating 
stronger nonflow at high \pt.

The AMPT-model calculations are in quantitative agreement with the kinematic 
dependence of \vtwo in these collision systems, indicating the breaking of 
flow factorization in this model. In midcentral to peripheral \heau collisions, 
below $\pt<1.5$~GeV/$c$, the AMPT model shows a clear separation between 
$v_2\{{\rm BF}\}$ and $v_2\{{\rm BB}\}$ unlike the experimental data, again 
indicating an overestimate of nonflow and decorrelation effects in this 
model.

As shown in Fig.~\ref{fig:pp_v2_pt}, the AMPT model \vtwo with the 
3$\times$2PC method also overestimates the experimental data in \pp 
collisions, similar to the comparison made for the peripheral \pau collision 
case. Again this overestimate may be attributable to the jet kinematics 
mismatch in the HIJING model used in AMPT~\cite{Lim:2019cys}. 
The large gap between \vtwo relative to the parton participant plane and 
that with the 3$\times$2PC method indicates nonflow is dominant in \pp 
collisions in the AMPT model.

\subsubsection{Multiplicity Dependence}

Figure~\ref{fig:v2_centrality} shows a comparison of AMPT \vtwo with 
the experimental results as a function of centrality. In 
$0.5<\pt<1$~GeV/$c$, the AMPT model \vtwo with the BB detector 
combination shows a flat trend in \pau collisions and slight decreasing 
trends in \dau and \heau collisions over the entire measured centrality 
ranges, which is inconsistent with the experimental data. In contrast, 
\vtwo with the BF detector combination shows an increasing trend towards the 
most peripheral collisions. For $2<\pt<2.5$~GeV/$c$, the AMPT model 
\vtwo with both detector combinations qualitatively captures the increasing 
trends in the experimental data.

Figure~\ref{fig:v2_vs_mult} shows a comparison of AMPT \vtwo with the 
experimental results as a function of $\frac{dN_{ch}}{d\eta}$. As seen in 
the centrality dependence of \vtwo, the AMPT model generally fails to 
reproduce the qualitative trends of $v_2\{{\rm BB}\}$ in $0.5<\pt<1$~GeV/$c$ 
while it captures the increasing trends of $v_2\{{\rm BB}\}$ in 
$2<\pt<2.5$~GeV/$c$ and $v_2\{{\rm BF}\}$ in both $0.5<\pt<1$~GeV/$c$ and 
$2<\pt<2.5$~GeV/$c$ towards smaller systems (and hence lower 
multiplicities). The AMPT simulations also show an increase of 
$v_2\{{\rm BB}\}$ and $v_2\{{\rm BF}\}$ for $0.5<\pt<1$~GeV/$c$ with 
increasing multiplicity above $\frac{dN_{ch}}{d\eta}=10$. This reflects the 
dominance of collective expansion at low-\pt in the AMPT model.

Finally, Fig.~\ref{fig:r_vs_mult} shows a comparison of the $R$ value 
calculated in AMPT with the experimental results as a function of 
$\frac{dN_{ch}}{d\eta}$. For $0.5<\pt<1$~GeV/$c$, the AMPT model simulations 
show an increasing trend in $R$ as $\frac{dN_{ch}}{d\eta}$ decreases, which 
is contradicted by the experimental data. However, the AMPT model is in 
agreement with the flow factorization seen in the experimental data at high 
$\frac{dN_{ch}}{d\eta}$. For $2<\pt<2.5$~GeV/$c$, the AMPT model 
calculations qualitatively capture the trends of the measured $R$ values.

\section{Summary}

In summary, measurements of azimuthal anisotropy \vtwo were presented as 
a function of \pt, centrality, and charged-particle multiplicity in MB 
\pp and noncentral \pau, \dau, and \heau collisions at \sqsntwo using 
the 3$\times$2PC method. The previous experimental findings that 
$v_2\{{\rm BF}\}>v_2\{{\rm BB}\}$ is also found in peripheral collisions 
in \pau, \dau, and \heau as well as in MB \pp collisions. This indicates 
smaller nonflow contribution in the BB combination and much more 
substantial nonflow contribution in the BF combination, in concurrence 
with the conclusions of Refs.~\cite{PHENIX:2021bxz,Nagle:2021rep}. The 
possible contributions to these \vtwo from the nonflow between the 
backward detectors and longitudinal decorrelation effects between the 
backward and forward detectors are under 
discussion~\cite{PHENIX:2021bxz, Nagle:2021rep} towards precise 
quantification of these effects. The kinematic dependence of \vtwo is 
quantified as the ratio $R$ of \vtwo between the two detector 
combinations as a function of $\frac{dN_{ch}}{d\eta}$ for $0.5<\pt<1$ 
and $2<\pt<2.5$~GeV/$c$. The different trend of $R$ between these \pt 
selections suggests strong \pt dependence of nonflow effects. The AMPT 
model calculations can quantitatively describe the experimental 
measurements only in most-central to midcentral \dau and \heau 
collisions, and it systematically overestimates in \pau and \pp 
collisions, indicating an unrealistically high nonflow contribution in 
AMPT. These measurements in various collision systems with different 
fractions of prehydrodynamization, nonflow, and decorrelation effects 
may serve as references for future unified models incorporating 
initial-state effects, prehydrodynamization dynamics, hydrodynamic 
expansion, and jets.


\begin{acknowledgments}

We thank the staff of the Collider-Accelerator and Physics
Departments at Brookhaven National Laboratory and the staff of
the other PHENIX participating institutions for their vital
contributions.  
We acknowledge support from the Office of Nuclear Physics in the
Office of Science of the Department of Energy,
the National Science Foundation,
Abilene Christian University Research Council,
Research Foundation of SUNY, and
Dean of the College of Arts and Sciences, Vanderbilt University
(USA),
Ministry of Education, Culture, Sports, Science, and Technology
and the Japan Society for the Promotion of Science (Japan),
Natural Science Foundation of China (People's Republic of China),
Croatian Science Foundation and
Ministry of Science and Education (Croatia),
Ministry of Education, Youth and Sports (Czech Republic),
Centre National de la Recherche Scientifique, Commissariat
{\`a} l'{\'E}nergie Atomique, and Institut National de Physique
Nucl{\'e}aire et de Physique des Particules (France),
J. Bolyai Research Scholarship, EFOP, the New National Excellence
Program ({\'U}NKP), NKFIH, and OTKA (Hungary),
Department of Atomic Energy and Department of Science and Technology
(India),
Israel Science Foundation (Israel),
Basic Science Research and SRC(CENuM) Programs through NRF
funded by the Ministry of Education and the Ministry of
Science and ICT (Korea),
Ministry of Education and Science, Russian Academy of Sciences,
Federal Agency of Atomic Energy (Russia),
VR and Wallenberg Foundation (Sweden),
University of Zambia, the Government of the Republic of Zambia (Zambia),
the U.S. Civilian Research and Development Foundation for the
Independent States of the Former Soviet Union,
the Hungarian American Enterprise Scholarship Fund,
the US-Hungarian Fulbright Foundation,
and the US-Israel Binational Science Foundation.

\end{acknowledgments}


%
 
\end{document}